\documentclass[12pt]{elsarticle}

\usepackage{graphicx}
\usepackage{multirow}
\usepackage{amssymb}

\journal{}

\begin{document}

\begin{frontmatter}


\title{Forecasting Bitcoin closing price series using linear regression and neural networks models}

\author{Nicola Uras
}
\ead{nicola.uras@unica.it}

\author{Lodovica Marchesi\corref{cor1}
}
\ead{lodovica.marchesi@unica.it}

\author{Michele Marchesi
}
\ead{marchesi@unica.it}

\author{Roberto Tonelli
}
\ead{roberto.tonelli@dsf.unica.it}
\cortext[cor1]{Corresponding author}




\address{Department of Mathematics and Computer Science \\
University of Cagliari \\
Cagliari, Italy }

\begin{abstract}
This paper studies how to forecast daily closing price series of Bitcoin, using data on prices and volumes of prior days. Bitcoin price behaviour is still largely unexplored, presenting new opportunities. We compared our results with two modern works on Bitcoin prices forecasting and with a well-known recent paper that uses Intel, National Bank shares and Microsoft daily NASDAQ closing prices spanning a 3-year interval. We followed different approaches in parallel, implementing both statistical techniques and machine learning algorithms. The \textit{SLR} model for univariate series forecast uses only closing prices, whereas the \textit{MLR} model for multivariate series uses both price and volume data. We applied the \textit{ADF}-Test to these series, which resulted to be indistinguishable from a random walk. We also used two artificial neural networks: \textit{MLP} and \textit{LSTM}. We then partitioned the dataset into shorter sequences, representing different price "regimes", obtaining best result using more than one previous price, thus confirming our regime hypothesis. All the models were evaluated in terms of \textit{MAPE} and \textit{relativeRMSE}. They performed well, and were overall better than those obtained in the benchmarks. Based on the results, it was possible to demonstrate the efficacy of the proposed methodology and its contribution to the state-of-the-art.
\end{abstract}

\begin{keyword}
Blockchain \sep Bitcoin \sep Time Series \sep Forecasting \sep Regression \sep Machine Learning \sep Neural Networks



\end{keyword}

\end{frontmatter}



\section{Introduction}
Bitcoin is the world's most valuable cryptocurrency, a form of electronic cash, invented by an unknown person or group of people using the pseudonym Satoshi Nakamoto \cite{A}, whose network of nodes was started in 2009. Although the system was introduced in 2009, its actual use began to grow only from 2013. Therefore, Bitcoin is a new entry in currency markets, though it is officially considered as a commodity rather than a currency, and its price behaviour is still largely unexplored, presenting new opportunities for researchers and economists to highlight similarities and differences with standard financial currencies, also in view of its very different nature with respect to more traditional currencies or commodities. The price volatility of Bitcoin is far greater than that of fiat currencies \cite{B}, providing significant potential in comparison to mature financial markets \cite{C} \cite{CoccoIEEEaccess} \cite{CoccoFI}.
According to \textit{coinmarketcap} website, one of the most popular sites that provides almost real-time data on the listing of the various cryptocurrencies in global exchanges, on May 2019 Bitcoin market capitalization value is valued at approximately 105 billion of USD. Hence, forecasting Bitcoin price has also great implications both for investors and traders.
Even if the number of bitcoin price forecasting studies is increasing, it still remains limited \cite{O}.
\vspace{3mm}

In this work, we approach the forecast of daily closing price series of the Bitcoin cryptocurrency using data on prices and volumes of prior days, 
and compare our results with three well-known recent papers, two dealing with Bitcoin prices forecasting using other approaches, 
and another one forecasting Intel, National Bank shares and Microsoft daily NASDAQ prices.    

The first paper we compare to, tries to predict three of the most challenging stock market time series data from NASDAQ historical quotes, namely Intel, National Bank shares and Microsoft daily closed (last) stock price, using a model based on chaotic mapping, firefly algorithm, and Support Vector Regression (SVR) \cite{G}.
In the second one Mallqui and Fernandes used different machine learning techniques such as Artificial Neural Networks (ANN) and Support Vector Machines (SVM) to predict, among other things, closing prices of Bitcoin \cite{O}.
The third paper we consider in our work proposes a two stage fusion approach to forecast stock market index. The first stage involves SVR. The second stage uses ANN, Random Forest (RF) and SVR \cite{N}.
In this work we forecast daily closing price series of Bitcoin cryptocurrency using data of prior days 
following different approaches in parallel, implementing both statistical techniques and machine learning algorithms. We tested the chosen algorithms on two datasets: the first consisting only of the closing prices of the previous days; the second adding the volume data. Since Bitcoin exchanges are open 24/7, the closing price reported on \textit{coinmarketcap} we used, refers to the price at 11:59 PM UTC of any given day. The implemented algorithms are Simple Linear Regression (SLR) model for univariate series forecast, using only closing prices; a Multiple Linear Regression (MLR) model for multivariate series, using both price and volume data; a Multilayer Perceptron and a Long Short-Term Memory neural networks tested using both the datasets.
The first step consisted in a statistical analysis of the overall series. From this analysis it turned out that the entire series are not distinguishable from a random walk. If the series were truly random walks, it would not be possible to make any forecasts.
Since we are interested in prices and not in price variations, we avoided the time series differencing technique by introducing and using the novel presented approach.
Therefore, each time series was segmented in shorter overlapping sequences in order to find shorter time regimes that do not resemble a random walk so that they can be easily modeled. Afterwards, we run all the algorithms again on the partitioned dataset.

The reminder of this paper is organized as follows. Section 2 presents the methodology, briefly describing the data, their pre-processing, and finally the models used. Section 3 presents and discuss the results. Section 4 concludes the paper.

\section{Literature Review}
Over the years many algorithms have been developed for forecasting time series in stock markets. The most widely adopted are based on the analysis of past market movements \cite{D}. Among the others, Armano et al. proposed a prediction system using a combination of genetic and neural approaches, having as inputs technical analysis factors that are combined with daily prices \cite{E}. Enke and Mehdiyev discussed a hybrid prediction model that combines differential evolution-based fuzzy clustering with a fuzzy inference neural network for performing an index level forecast \cite{F}. Kazem et al. presented a forecasting model based on chaotic mapping, firefly algorithm, and support vector regression (SVR) to predict stock market prices \cite{G}. Unlike other widely studied time series, still very few researches have focused on bitcoin price prediction. A recent exploration tries to ascertain with what accuracy the direction of Bitcoin price in USD can be predicted using machine learning algorithms like LSTM (Long short-term memory) and RNN (Recurrent Neural Network) \cite{H}. Naimy and Hayek (2018) tried to forecast the volatility of the Bitcoin/USD exchange rate using GARCH (Generalized AutoRegressive Conditional Heteroscedasticity) models \cite{I}. D. U. Sutiksno et al. studied and applied $\alpha$-Sutte indicator and Arima (Autoregressive Integrated Moving Average) methods to forecast historical data of Bitcoin \cite{L}. Stocchi and Marchesi proposed the use of Fast Wavelet Transform to forecast Bitcoin prices \cite{M}.
Steve Y. Yang et al. examined a few complexity measures of the Bitcoin transaction flow networks, and modeled the joint dynamic relationship between these complexity measures and Bitcoin market variables such as return and volatility \cite{Aggiunta1}.
Nashirah A. Bakar and S. Rosbi presented a forecasting Bitcoin exchange rate model in high volatility environment, using autoregressive integrated moving average (ARIMA) algorithms \cite{Aggiunta2}.
L. Catania et al. studied the predictability of cryptocurrencies time series, comparing several alternative univariate and multivariate models in point and density forecasting of four of the most capitalized series: Bitcoin, Litecoin, Ripple and Ethereum, using univariate Dynamic Linear Models and several multivariate Vector Autoregressive models with different forms of time variation \cite{Aggiunta3}.
Nhi N.Y. Vo and G. Xu used knowledge of statistics for financial time series and machine learning to fit the parametric distribution and model and forecast the volatility of Bitcoin returns, and analyze its correlation to other financial market indicators \cite{Aggiunta4}.
Other approaches try to predict stock market index using fusion of machine learning techniques \cite{N}.

\section{Methodology}
In this section we first introduce some notions on time series analysis, which helped us to take the operational decisions about the algorithms we used and to better understand the results presented in the following. Then, we present the dataset we used, including its pre-processing analysis. Finally we introduce our proposed algorithms with the metrics employed to evaluate their performance and the statistical tools we adopted.

\subsection{Time Series Analysis}
\subsubsection{Time Series Components}
Any time series is supposed to consist of three systematic components that can be described and modelled. These are 'base level', 'trend' and 'seasonality', plus one non-systematic component called 'noise'.
The base level is defined as the average value in the series. A trend is observed when there is an increasing or decreasing slope in the time series. Seasonality is observed when there is a repeated pattern between regular intervals, due to seasonal factors. Noise represents the random variations in the series.
Every time series is a combination of these four components, where base level and noise always occur, whereas trend and seasonality are optional.
Depending on the nature of the trend and seasonality, a time series can be described as an additive or multiplicative model. This means that each observation in the series can be expressed as either a sum or a product of the components \cite{P}. 
An additive model is described by following the linear equation:
\begin{equation}
    y(t) = BaseLevel + Trend + Seasonality + Noise
\end{equation}


A multiplicative model is instead represented by the following non -linear equation: 
\begin{equation}
    y(t) = BaseLevel * Trend * Seasonality * Noise
\end{equation}

An additive model would be used when the variations around the trend does not vary with the level of the time series whereas a multiplicative model would be appropriate if the trend is proportional to the level of the time series.
This method of time series decomposition is called "classical decomposition" \cite{P}.
\vspace{0,3cm}

\subsubsection{Statistical Measures}
The statistical measures we calculated for each time series are the mean, labelled with $\mu$, the standard deviation $\sigma$ and the trimmed mean $\bar\mu$, obtained discarding a portion of data from both tails of the distribution.
The trimmed mean 
is less sensitive to outliers than the mean, but it still gives a reasonable estimate of central tendency and can be very helpful 
for time series with high volatility. 

\subsection{Collected data}
We tested our algorithms on four daily price series. Three of them are stock market series, all extracted from the 'Historical Data' available on \textit{yahoofinance} website; the fourth one is the Bitcoin price daily series, extracted from \textit{coinmarketcap} website.
\begin{itemize}
    \item Daily stock market prices for Microsoft Corporation (MSFT), from 9/12/2007 to 11/11/2011.
    \item Daily stock market prices for Intel Corporation (INTC), from 9/12/2007 to 11/11/2010.
    \item Daily stock market prices for National Bankshares Inc. (NKSH), from 6/27/2008 to 8/29/2011.
    \item Daily Bitcoin price series, from 15/11/2015 to 11/08/2018.
\end{itemize}

We state once more that we choose these price series and the related 
time intervals as benchmark to compare our results with well known literature results obtained by using other methods. 
%
Specifically, we have chosen for the stock market series the same 
time intervals chosen in \cite{G}. 
The choice of Bitcoin as criptocurrency is quite natural since 
it represents about 58$\%$ of the Total Market Capitalization.
It is worth noting that, because of the recent birth of Bitcoin and its recent actual growth (from 2013 on), it was not possible to employ the same time interval for all price series included in the study.

The dataset was divided into two sets, a training part and a testing part. After some empirical test the partition of the data which lead us to optimal solutions was 80$\%$ of the daily data for the training dataset and the remaining for the testing dataset.

\subsection{Data pre-processing}
For both models we prepared our dataset in order to have a set of inputs ($X$) and  outputs ($Y$) with temporal dependence. 
We performed a one-step ahead forecast: our output $Y$ is the value from the next (future) point of time while the inputs $X$ are one or several values from the past, i.e. the so called \textit{lagged} values. 
From now on we identify the number of used lagged values with the \textit{lag} parameter.
In the Linear Regression and \textit{Univariate} LSTM models the dataset includes only the daily closing price series, hence there is only one single $lag$ parameter for the \textit{close} feature. On the contrary, in the Multiple Linear Regression and \textit{Multivariate} LSTM models the dataset includes both \textit{close} and \textit{volume (USD)} series, hence we use two different $lag$ parameters, one for the \textit{close} and one for  the \textit{volume} feature.
In both cases, we attempted to optimize the predictive performance of the models by varying the \textit{lag} 
from $1$ to $10$.

\subsection{Univariate versus Multivariate Forecasting}
A univariate forecast consists of predicting time series made by observations belonging to a single feature recorded over time, in our case the closing price of the series considered. A multivariate forecast is a forecast in which the dataset consists of the observations of several features. In our case we used:
\begin{itemize}
    \item for BTC series all the features provided by \textit{coinmarketcap} website: Open, High, Low, Close, Volume.
    \item for MSFT, INTC, NKSH series all the features provided by \textit{yahoofinance} website: Date, Open, High, Low, Close, Volume.
\end{itemize}
We observed that adding features to the dataset did not lead to better predictions, but performance and results worsened. For this reason, we decided to use in the multivariate analysis only the \textit{close} and \textit{volume} features, that provided the best results.

\subsection{Statistical Analysis}
As a first step we carried out a statistical analysis in order to check for non-stationarity in the time series. We used the \textit{augmented Dickey-Fuller test} and \textit{autocorrelation plots} \cite{Q} \cite{R}.
To facilitate the understanding of this analysis we introduce the concept of \textit{unit root}.
A stochastic process with a \textit{unit root}  
is non-stationary, namely shows statistical properties that change over time, including mean, variance and covariance, and can cause problems in statistical inference involving time series models. A common process with \textit{unit root} is the \textit{random walk}. Often time series show some characteristics which makes them indistinguishable from a random walk.
The presence of such a process can be tested using a \textit{unit root} test.

The \textit{ADF} test is a statistical test that can be used to test for a \textit{unit root} in a univariate process, such as time series samples. 
The null hypothesis $H_0$ of the \textit{ADF} test is that there is a \textit{unit root}, with the alternative $H_a$ that there is no \textit{unit root}.
The most significant results provided by this test are the \textit{observed test statistic}, the Mackinnon's approximate \textit{p-value} and the \textit{critical values} at the $1\%$, $5\%$ and $10\%$ levels.

The test statistic is simply the value provided by the \textit{ADF} test for a given time series.
Once this value is computed it can be compared to the relevant critical value for the Dickey-Fuller Test.

Critical values, usually referred to as $\alpha$ levels, are an error rate defined in the hypothesis test. They give the probability to reject the null hypothesis $H_0$.
So if the observed test statistic is less than the critical value (keep in mind that ADF statistic values are always negative \cite{Q}), then the null hypothesis $H_0$ is rejected and no \textit{unit root} is present.

The \textit{p-value} is instead the probability to get a "more extreme" test statistic than the one observed, based on the assumed statistical hypothesis $H_0$, and its mathematical definition is shown in equation~\ref{eq:p-value}.
\begin{equation}
p_{value} = P\Big(t \geq t_{observed} 
\Big| \hspace{0.1cm} H_0\Big)
\label{eq:p-value}
\end{equation}

The \textit{p-value} is sometimes called \textit{significance}, actually meaning the closeness of the \textit{p-value} to zero: the lower the \textit{p-value}, the higher the significance.

In our analysis we performed this test using the \textit{adfuller()} function provided by the \textit{statsmodels} Python library, and we chose a \textit{significance level} of $5\%$.

Furthermore, the \textit{autocorrelation plot}, also known as \textit{correlogram}, allowed us to calculate the correlation between each observation and the observations at previous time steps, called \textit{lag values}.
In our case we employed the \textit{autocorrelation\_plot()} function provided by the python \textit{Pandas} library \cite{PANDAS}.

\subsection{Forecasting}
We decided to follow two different approaches: the first uses two well-known statistical methods: Linear Regression (LR) and Multiple Linear Regression (MLR). The second uses two very common neural networks (NN): Multilayer Perceptron (MLP) NN and Long Short-Term Memory (LSTM) NN. The reasons of this choices are explained below.

\subsubsection{Linear Regression and Multiple Linear Regression}
Linear regression is a linear approach for modelling the relationship between a dependent variable and one independent variable, represented by the main equation:

\begin{equation}
    y = b_0 + \vec{b}_1 \cdot \vec{x}_1,
\end{equation}


where $y$ and $\vec{x}_1$ are the dependent and the independent variable respectively, while $b_0$ is the intercept and $\vec{b}_1$ is the vector of slope coefficients.
In our case the components of the vector $\vec{x}_1$, our independent variable, are the values of the closing prices of the previous days. Therefore, $\vec{x}_1$ size is the value of the \textit{lag} parameter.
In our case $y$ represents the closing price to be predicted. 

This algorithm aims to find the curve that best fits the data, which best describes the relation between the dependent and independent variable.
The algorithm finds the best fitting line plotting all the possible trend lines through our data and for each of them calculates and stores the amount $(y-\bar{y})^{2}$, and then choose the one that minimizes the squared differences sum $\sum_{i} (y_i - \bar{y}_i)^{2}$, namely the line that minimizes the distance between the real points and those crossed by the line of best fit.

We then tried to forecast with multiple independent variables, adding to the \textit{close} price feature the observations of several features, including \textit{volume}, \textit{highest value} and \textit{lowest value} of the previous day.
These information were gained from the \textit{coinmarketcap} website.
In these cases we used a Multiple Linear Regression model (MLR). The MLR equation is:
\begin{equation}
    y = b_0 + \vec{b}_1\cdot \vec{x}_1+ ... + 
    \vec{b}_n\cdot \vec{x}_n = b_0 +  \sum_{i=1}^{n} \vec{b}_i \cdot \vec{x}_i
\end{equation}
where the index $i$ refers to a particular independent variable and $n$ is the dimension of the independent variables space.

We used the Linear and Multiple regression model of \textit{scikit learn} \cite{S}.
We decided to use this two models for several reasons: they are simple to write, use and understand, they are fast to compute, they are commonly used models and fit well to datasets with few features, like ours.
Their disadvantage is that they can model only linear relationships.

\subsubsection{Multilayer Perceptron}
A multilayer perceptron (MLP) is a feedforward artificial neural network that generates a set of outputs from a set of inputs. It consists of at least three layers of neurons: an input layer, a hidden layer and an output layer. Each neuron, apart from the input ones, has a nonlinear activation function. MLP uses backpropagation for training the network.
In our model we keep the structure as simple as possible, with a single hidden layer. Our inputs are the closing prices of the previous days, where the number of values considered depends on the \textit{lag} parameter. The output is the forecast price. The optimal number of neurons were found by optimizing the network architecture on the number of neurons itself, varying it in an interval between 5 and 100. We used the Python Keras library \cite{KERAS}.

\subsubsection{LSTM Networks}
Long Short-Term Memory networks are nothing more than a prominent variations of Recurrent Neural Network (RNN).
RNN's are a class of artificial neural network with a specific architecture oriented at recognizing patterns in sequences of data of various kinds: texts, genomes, handwriting, the spoken word, or numerical time series data emanating from sensors, markets or other sources \cite{LSTM}. Simple recurrent neural networks are proven to perform well only for short-term memory and are unable to capture long-term dependencies in a sequence. On the contrary, LSTM networks are a special kind of RNN, able at learning long-term dependencies.
The model is organized in cells which include several operations. LSTM hold an internal state variable, which is passed from one cell to another and modified by Operation Gates (forget gate, input gate, output gate). These gates control how much of the internal state is passed to the output and work in a similar way to other gates. These three gates have independent weights and biases, hence the network will learn how much of the past output and of the current input to retain and how much of the internal state to send out to the output.

In our case the inputs are the closing prices of the previous days and the number of values considered depends on the \textit{lag} parameter. The output is the forecast price.
We used the Keras framework for deep learning. Our model consists of one stacked LSTM layer with 64 units each and the densely connected output layer with one neuron. We used Adam optimizer and MSE (mean squared error) as a loss. We performed several experiments and found that the optimal number of epochs and batch size are 600 and 72 respectively for forecasting our Bitcoin daily closing prices. We set shuffle=False because we did not want to shuffle time series data.

\subsection{Time Regimes}
The time series considered are found to be indistinguishable from a random walk. This peculiarity is common for time series of financial markets, and in our case is confirmed by the predictions of the models, in which the best result is obtained considering only the price of the previous day. 
The purpose is to find an approach that allow us to avoid time series differencing technique, in view of the fact that we are interested in prices and not in price variations represented by integrated series of $d$-order.
For this reason, each time series was segmented into short partially overlapping sequences, in order to find if shorter time regimes are present, where the series do not resemble a random walk.
Finally, to continue with the forecasting procedure, a train and a test set were identified within each time regime.

For each regime we always sampled $200$ observations. The beginning of the next regime is obtained with a shift of $120$ points from the previous one. Thus, every regime is $200$ points wide and has $80$ points in common with the following one.
Since the time series considered have different lengths, the partition in regimes has generated:
\begin{itemize}
    \item Bitcoin: 7 regimes
    \item Microsoft: 8 regimes
    \item Intel and National Bankshares: 5 regimes
\end{itemize}
From a mathematical point of view, the used approach can be described as follows.

Let us target a vector $\overrightarrow{OA}$ along the $t$ axis, with length $200$.
This vector is identified by the points $O(1, 0)$, $A(a, 0)\equiv{(200, 0)}$.
The length of this vector represents the width of each time regime.

Let $\overrightarrow{OH}$ be a fixed translation vector along the $t$ axis, identified by the points $O(1, 0)$ and $H(h, 0)\equiv{(120, 0)}$. The length of $\overrightarrow{OH}$ represents the translation size.

For the sake of simplicity, let us label the $\overrightarrow{OA}$ and $\overrightarrow{OH}$ vectors with $\vec{A}$ and $\vec{H}$.

Let $\vec{A'}$ be the vector $\vec{A}$ shifted by $\vec{H}$ and $\vec{A}^{n}$ the vector $\vec{A}$ shifted by $n$ times $\vec{H}$.

Therefore, the vector that identifies the $n$th sequence to be sampled along the series is given by:
\begin{equation}
    \label{eq:vectorA}
    \vec{A}^{n} = \vec{A} + n\vec{H}
\end{equation}
where $n \in \big[0, \frac{D-A}{h}\big]$, being $D$ the dimension of the sampling space, $A$ the time regimes width and $h$ the translation size.

So the $n$th time regime is given by:
\begin{equation}
\label{eq:regime}
R^{n} = f\big(\vec{A}^{n}\big) = f\big(\vec{A} + n\vec{H}\big)
\end{equation}


where $f$ is the function that maps the values along the $t$ axis (dates) to the respective regimes $y$ values (actual prices).

\subsection{Performance Measures}
To evaluate the effectiveness of different approaches, we used the \textit{relative} Root Mean Square Error (rRMSE) and the Mean Absolute Percentage Error (MAPE), 
defined respectively as:
\begin{equation}
\label{eq:rRMSE}
    relativeRMSE = \sqrt{\frac{1}{N}\sum_{i=1}^{N}\Big(\frac{y_i - f_i}{y_i}\Big)^{2}}
\end{equation}
\begin{equation}
\label{eq:MAPE}
MAPE =   \frac{1}{N}\sum_{i=1}^{N}\Big|\frac{y_i - f_i}{y_i}\Big|
\end{equation}

In both formulas $y_i$ and $f_i$ represent the actual and forecast values, and $N$ is the number of forecasting periods.
These are scale free performance measures, so that they are well appropriate to compare model performance results across series with different orders of magnitude, as in our study.

\section{Results}

\subsection{Time Series Analysis}
\begin{figure*}[htbp]
\centering
\includegraphics[width=1\textwidth]{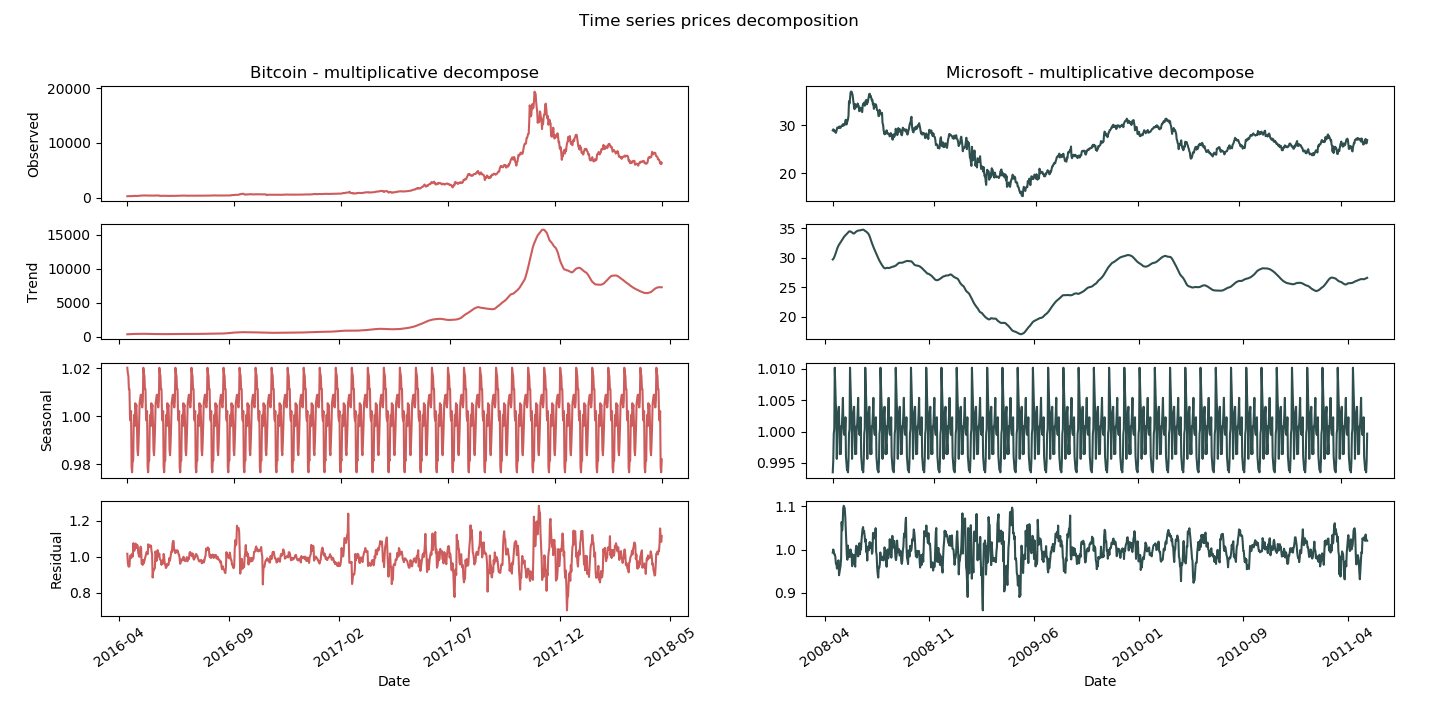}
\caption{Decomposition of Bitcoin and Microsoft Time Series}
\label{fig:figureOne}
\end{figure*}

In figure~\ref{fig:figureOne} we report the decomposition of Bitcoin and Microsoft time series, for comparison purposes, as obtained 
using the \textit{seasonal\_decompose()} method, provided by the Python \textit{statsmodels} library \cite{STATSMODELS}. 

The \textit{seasonal\_decompose()} method requires to specify whether the model is additive or multiplicative.
In the Bitcoin time series, the trend of increase at the beginning is almost absent (from around 2016-04 to 2017-02); in later years, the frequency and the amplitude of the cycle appears to change over time.
The Microsoft time series shows a non-linear seasonality over the whole period, with frequency and amplitude of the cycles changing over time.
These considerations suggest that the model is multiplicative.
Furthermore, if we look at the residuals, they look quite random, in agreement with their definitions.
The Bitcoin residuals are likewise meaningful, showing periods of high variability in the later years of the series.

\begin{figure*}[htbp]
\centering
\includegraphics[width=1\textwidth]{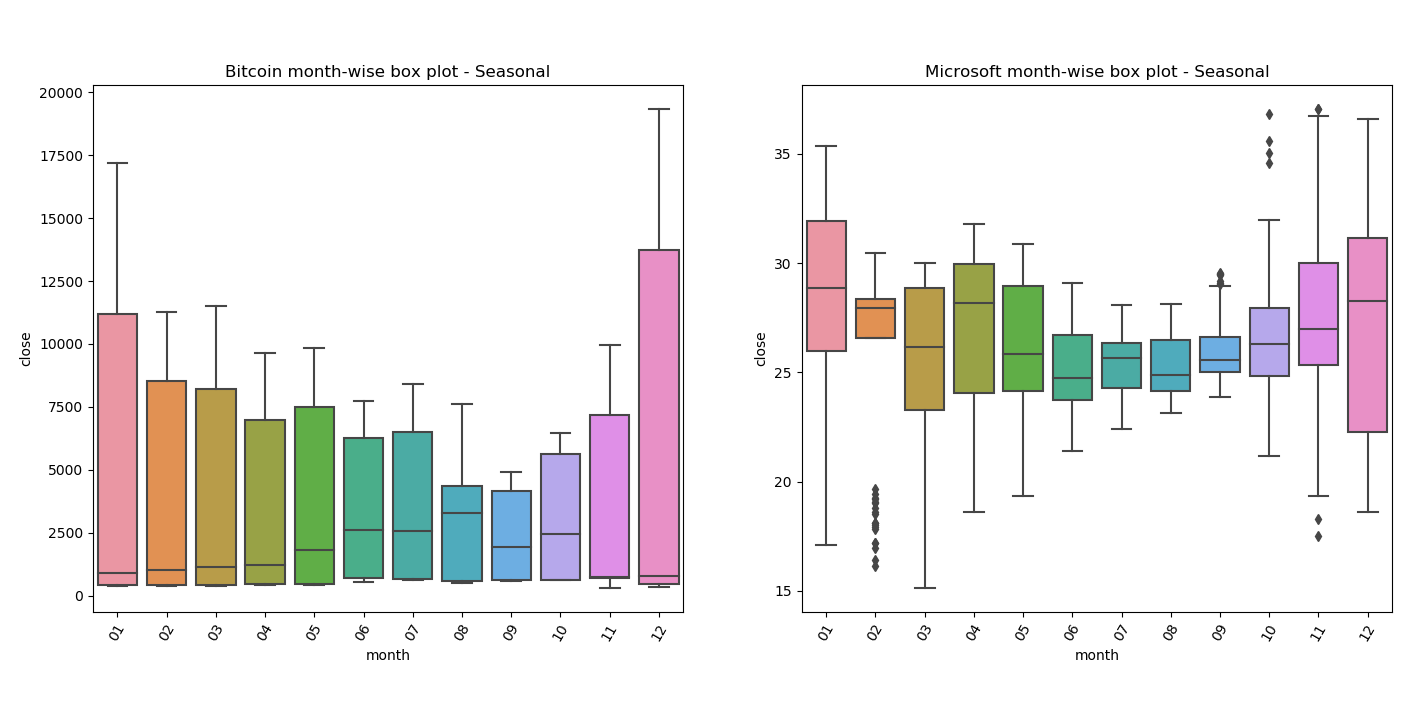}
\caption{Seasonality of Bitcoin and Microsoft Time Series}
\label{fig:figureTwo}
\end{figure*}

It is also possible to group the data at seasonal intervals, observing how the values are distributed and how they evolve over time. In our work we grouped the data of the same month over the years we considered.
This is achieved with the 'Box plot' of month-wide distribution, shown in figure~\ref{fig:figureTwo}.

The Box plot is a standardized way of displaying the distribution of data based on five numbers summary: minimum, first quartile, median, third quartile and maximum.
The box of the plot is a rectangle which encloses the middle half of the sample, with an end at each quartile. The length of the box is thus the inter-quartile range of the sample. The other dimension of the box has no meaning. A line is drawn across the box at the sample median.
Whiskers sprout from the two ends of the box 
defining the outliers range.
The box length gives an indication of the sample variability, and for the Bitcoin samples shows a large variance, in particular from December 2017 to March 2018.
The line crossing the box shows where the sample is centred, i.e. the median.
The position of the box in its whiskers and the position of the line in the box also tell us whether the sample is symmetric or skewed, either to the right or to the left.
The plot shows that the Bitcoin monthly samples are therefore skewed to the right. The top whiskers is much longer than the bottom whiskers and the median is gravitating towards the bottom of the box. This is due to the very high prices that Bitcoin reached throughout the period between 2017 and 2018. These large values tend to skew the sample statistics.
In Microsoft, an alternation between samples skewed to the left and samples skewed to the right occurs, except for the sample of October that shows a symmetric distribution. Lack of symmetry entails one tail being longer than the other, distinguishing between heavy-tailed or light-tailed populations. In the Bitcoin case we can state that the majority of the samples are left skewed populations with short tails.
Microsoft shows an alternation between heavy-tailed and light-tailed distributions. We can see that some Microsoft samples, particularly those with long tails, present outliers, representing anomalous values.
This is due to the fact that heavy tailed distributions tend to have many outliers with very high values. The heavier the tail, the larger the probability that you will get one or more disproportionate values in a sample.

\begin{table}
\begin{center}
\caption{Time Series Statistical Measures}
\label{table:tableOne}
\begin{tabular}{|p{1,5cm}|p{1,5cm}|p{1,5cm}|p{1,5cm}|}
\hline
\textbf{Series} & \boldmath{$\mu$} & \boldmath{$\sigma$} & \boldmath{$\bar\mu$} \\
\hline
\textit{BTC} & 3622,5 & 4083,2 & 2913,9 \\
\hline
\textit{MSFT} & 26,2 & 3,9 & 26,3 \\
\hline
\textit{INTC} & 19,9 & 3,6 & 19,9\\
\hline
\textit{NKSH} & 24,3 & 3,9 & 24,5\\
\hline
\end{tabular}
\end{center}
\end{table}

\begin{table}[ht]
\caption{Regimes Statistical Measures}
\label{table:tableTwo}
\begin{center}
\begin{tabular}{ |p{1cm}|p{1cm}|p{1,2cm}|p{1,2cm}|p{1,2cm}| } 
\hline
\textbf{Series} & \textbf{h} & \boldmath{$\mu$} & \boldmath{$\sigma$} & \boldmath{$\bar\mu$} \\
  \hline
{BTC} & 0 & 419,0 & 39,5 & 420,9 \\
  & 120 & 549,9 & 97,2 & 548,6 \\ 
  & 240 & 705,6 & 120,9 & 691,3 \\ 
  & 360 & 1106,5 & 356,1 & 1046,9 \\ 
  & 480 & 2483,9 & 1116,4 & 2415,4 \\ 
  & 600 & 7384,1 & 4719,8 & 6829,7 \\ 
  & 720 & 10278,8 & 3002,7 & 9904,6 \\ 
  \hline
  {MSFT} & 0 & 30,7 & 2,8 & 30,5 \\ 
  & 120 & 26,1 & 3,2 & 26,4 \\ 
  & 240 & 20,6 & 3,9 & 20,4 \\ 
  & 360 & 22,8 & 3,8 & 22,8 \\ 
  & 480 & 28,2 & 2,3 & 28,4 \\ 
  & 600 & 26,8 & 2,2 & 26,7 \\ 
  & 720 & 26,1 & 1,3 & 26,1 \\ 
  & 840 & 26,0 & 1,2 & 26,0 \\
  \hline
  {INTC} & 0 & 23,5 & 2,4 & 23,5 \\ 
  & 120 & 20,0 & 3,6 & 20,3 \\ 
  & 240 & 15,4 & 2,3 & 15,1 \\ 
  & 360 & 17,3 & 2,3 & 17,4 \\ 
  & 480 & 20,6 & 1,4 & 20,4 \\
  \hline
  {NKSH} & 0 & 18,5 & 0,9 & 18,5 \\ 
  & 120 & 22,2 & 3,0 & 22,2 \\ 
  & 240 & 26,5 & 1,4 & 26,5 \\ 
  & 360 & 25,9 & 1,9 & 26,0 \\ 
  & 480 & 26,5 & 2,5 & 26,3 \\
  \hline
\end{tabular}
\end{center}
\end{table}

Tables~\ref{table:tableOne} and~\ref{table:tableTwo} show the statistics calculated for each time series and for each short time regime. The unit of measurement of the values in the tables is the US dollar (\$). In table~\ref{table:tableOne} we can observe that the only series for which the trimmed mean, obtained with \textit{trim\_mean()} method provided by the Python \textit{scipy} library \cite{SCIPY}, with a cut-off percentage of $10\%$, is significantly different from the mean is BTC. In particular the trimmed mean decreased. This is due to the fact that the BTC, for a long period of time, registered a large price increment and this implies a shift of the mean to the right (i.e. to highest prices). This confirms that BTC distribution is right-skewed.
Table~\ref{table:tableTwo} shows that stock market series time regimes present a lower $\sigma$ than BTC ones, namely that BTC distribution has higher variance.

\begin{figure}[h]
\centering
\includegraphics[width=1\linewidth]{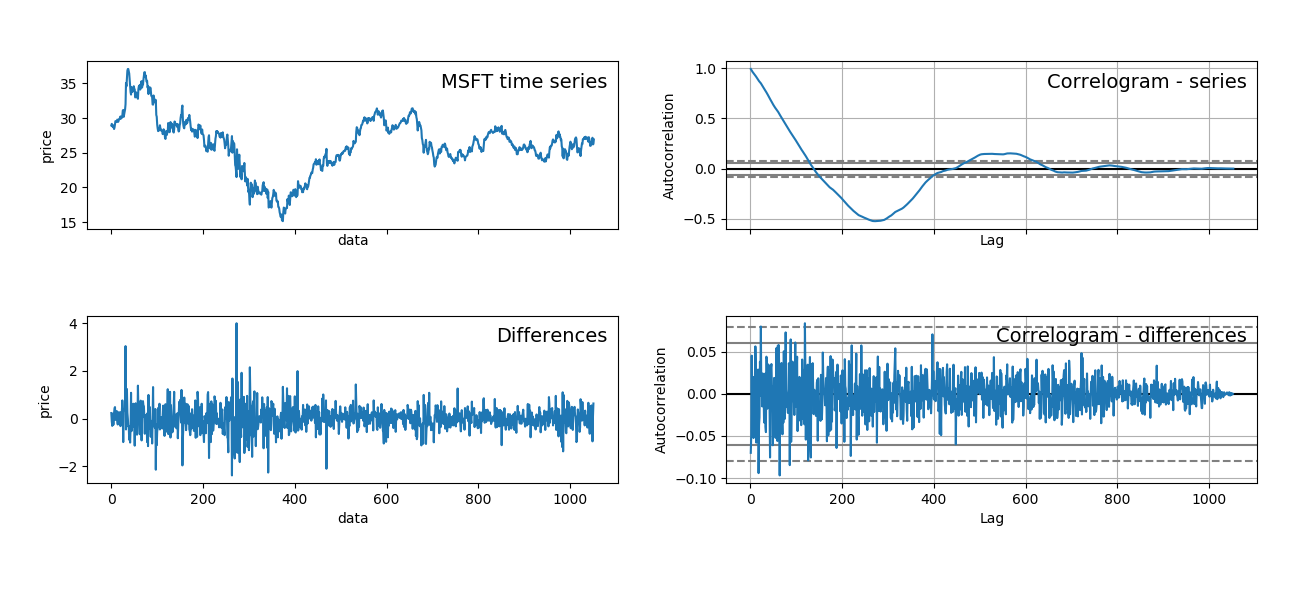}
\caption{Microsoft time series autocorrelation plots}
\label{fig:figureThree}
\end{figure}

\begin{figure*}[htbp]
\centering
\includegraphics[width=1\textwidth]{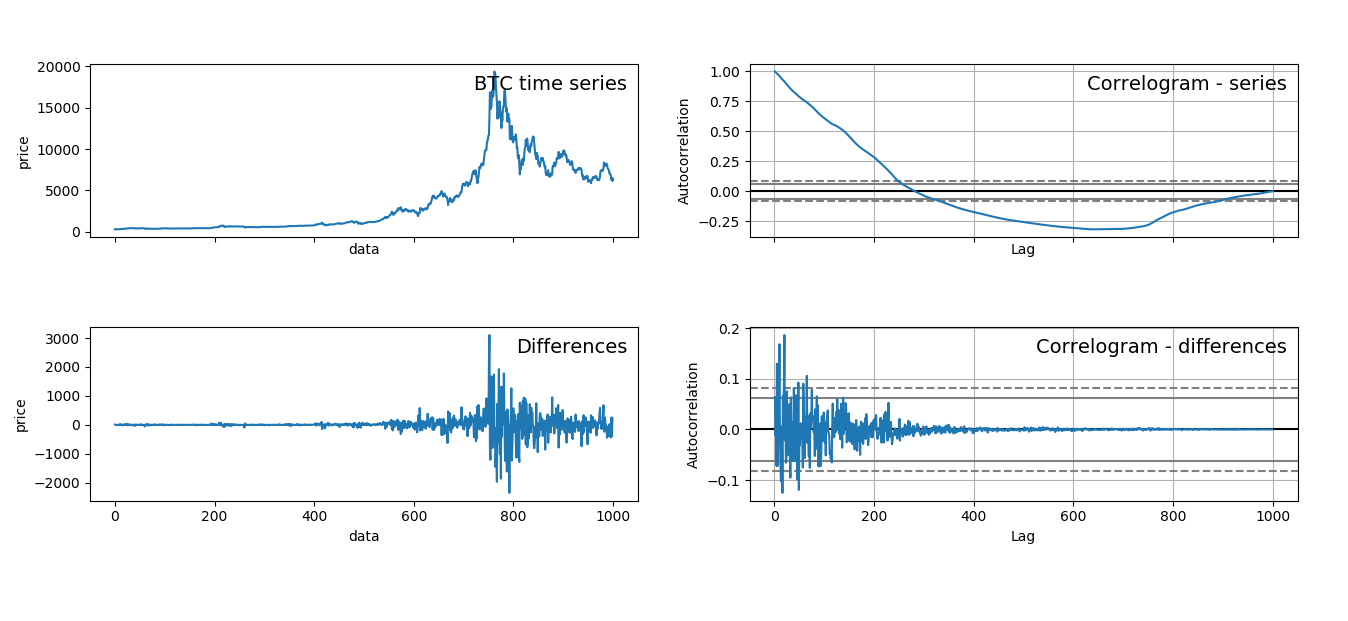}
\caption{Bitcoin time series autocorrelation plots}
\label{fig:figureFour}
\end{figure*}

Figures~\ref{fig:figureThree} and~\ref{fig:figureFour} show the autocorrelation plots of BTC and MSFT series. The others stock market series are not presented because they show the same features of the MSFT series.
Both autocorrelation plots show a strong autocorrelation between the current price and the closest previous observations and a linear fall-off from there to the first few hundred lag values.
We then tried to make the series stationary by taking the \textit{first difference}.
The autocorrelation plots of the 'differences series' show no significant relationship between the lagged observations. All correlations are small, close to zero and below the 95\% and 99\% confidence levels.

\begin{table}
\begin{center}
\caption{Augmented Dickey-Fuller test results}
\label{table:tableThree}
\begin{tabular}{|p{2cm}|p{3cm}|p{2cm}|}
\hline
\textbf{Series} & \textbf{ADF statistic} & \textbf{p-value} \\
\hline
\textit{BTC} & -1,61 & 0,47  \\
\hline
\textit{MSFT} & -1,98 & 0,29  \\
\hline
\textit{INTC} & -1,98 & 0,29  \\
\hline
\textit{NKSH} & -2,10 & 0,25  \\
\hline
\end{tabular}
\end{center}
\end{table}

As regards the \textit{augmented Dickey-Fuller} results, shown in table~\ref{table:tableThree}, looking at the observed \textit{test statistics}, we can state that all the series follows a unit root process.
We remind that the null hypothesis $H_0$ of the \textit{ADF} test is that there is a \textit{unit root}.
In particular, all the observed \textit{test statistics} are greater than those associated to all significance levels.
This implies that we can not reject the null hypothesis $H_0$, but does not imply that the null hypothesis is true.

Observing the \textit{p-values}, we notice that for the stock market series we have a low probability to get a "more extreme" test statistic than the one observed under the null hypothesis $H_0$. Precisely, for both \textit{MSFT} and \textit{INTC} we got a probability of $29\%$, for \textit{NKSH} a probability of $25\%$.
The Bitcoin series shows instead a slightly larger probability of $47\%$.
We conclude that $H_0$ can not be rejected and so each time series present a \textit{unit root} process.
We conclude that all the considered series show the statistical characteristics typical of a \textit{random walk}.

\subsection{Time Series Forecasting}
Table~\ref{table:tableFour} and~\ref{table:tableFive} show the results, in terms of MAPE and rRMSE, obtained with the different algorithms applied to the entire series. From now on, let us label the closing and the volume features \textit{lag} parameters with $k_p$ and $k_v$ respectively.
In particular,  table~\ref{table:tableFour} reports 
the best results obtained using the \textit{Linear Regression} algorithm for univariate series forecast, using only closing prices, and the \textit{Multiple Linear Regression} model for multivariate series, using both price and volume data.

Table~\ref{table:tableFive} shows the best results obtained with the \textit{LSTM} neural network, distinguishing between \textit{univariate LSTM}, using only closing prices, and \textit{multivariate LSTM}, using both price and volume data. 

Small values of the $MAPE$ and $rRMSE$ evaluation metrics suggest accurate predictions and good performance of the considered model.

From the analysis of the series in their totality, it appears that linear models outperforms neural networks.
However, for both models, best results are obtained for a \textit{lag} of $1$, except for the $NKSH$ time series, thus confirming our hypothesis that the series are indistinguishable from a random walk.

In order to perform the time series forecasting, we also implemented a \textit{Multi-Layer Perceptron} model. Since the \textit{LSTM} network outperforms the \textit{MLP} one, we decided to show only the \textit{LSTM} results. This is probably due to the particular architecture of the LSTM network, that is able to capture long-term dependencies in a sequence.

It should be noted that better predictions are obtained for stock market series rather than for the Bitcoin one. This is probably due to the high price fluctuations that Bitcoin has suffered during the investigated time interval. This is confirmed by the statistics shown in table ~\ref{table:tableOne}.

It must be noted that the addition of the \textit{volume} feature to the dataset does not improve the predictions. 

\begin{table}[h!]
\caption{Linear (left) and Multiple Linear Regression (right) results}
\label{table:tableFour}
\begin{center}
\begin{tabular}{ |p{1.1cm}|p{1.4cm}|p{1.5cm}|p{0.4cm}|p{1.4cm}|p{1.5cm}|p{0.4cm}|p{0.3cm}| } 
\hline
  \hline
  \textbf{Series} & \textbf{MAPE} & \textbf{rRMSE} & \boldmath{$k_p$} & \textbf{MAPE} & \textbf{rRMSE} & \boldmath{$k_p$} & \boldmath{$k_v$} \\
  \hline 
  BTC & 0,030 & 0,046 & 1 & 0,030 & 0,046 & 1 & 1\\
  MSFT & 0,010 & 0,015 & 1 & 0,010 & 0,015 & 1 & 1\\
  INTC & 0,010 & 0,017 & 1 & 0,010 & 0,017 & 1 & 1\\
  NKSH & 0,010 & 0,020 & 12 & 0,010 & 0,019 & 12 & 1\\
  \hline
\end{tabular}
\end{center}
\end{table}

\begin{table}[h!]
\caption{Univariate (left) and Multivariate (right) LSTM results}
\label{table:tableFive}
\begin{center}
\begin{tabular}{ |p{1.2cm}|p{1.4cm}|p{1.5cm}|p{0.3cm}|p{1.4cm}|p{1.5cm}|p{0.3cm}|p{0.3cm}| }
\hline
  \hline
  \textbf{Series} & \textbf{MAPE} & \textbf{rRMSE} & \boldmath{$k_p$} & \textbf{MAPE} & \textbf{rRMSE} & \boldmath{$k_p$} & \boldmath{$k_v$} \\
  \hline  
  BTC & 0,036 & 0,049 & 1 & 0,036 & 0,048 & 1 & 1\\
  MSFT & 0,012 & 0,015 & 1 & 0,012 & 0,015 & 1 & 1\\
  INTC & 0,014 & 0,017 & 2 & 0,014 & 0,017 & 2 & 2\\
  NKSH & 0,014 & 0,020 & 5 & 0,014 & 0,020 & 5 & 5\\
  \hline
\end{tabular}
\end{center}
\end{table}

In order to perform prices forecast we changed the approach and decided to split the time series analysis using shorter time 
windows of 200 points, shifting the windows by 120 points, as in Table~\ref{table:tableTwo}, with the aim of finding local time regimes where the series do not follow the global random walk pattern. 

Table~\ref{table:tableEight} and~\ref{table:tableNine} show the results obtained with our 
approach of partitioning the series into shorter sequences. Let us label the moving step forward with $h$.
Particularly, in table~\ref{table:tableEight} are presented the best results obtained using the \textit{Linear Regression} algorithm for univariate series forecast, using only closing prices, and the \textit{Multiple Linear Regression} model for multivariate series, using both price and volume data.
Table~\ref{table:tableNine} shows the best results obtained with the \textit{LSTM} neural network, distinguishing between \textit{univariate LSTM}, using only closing prices, and \textit{multivariate LSTM}, using both price and volume data.
For the sake of brevity we only show the best results obtained on a specific time window defined by the $h$ value reported in Tabs.~\ref{table:tableEight} and~\ref{table:tableNine} but all other analyzed windows (other $h$ values) provide  
MAPE and rRMSE lower than those obtained in the benchmark papers. In particular, we obtained outperforming MAPE and rRMSE for the Bitcoin time series also with respect to the financial ones 
(eg. for h = 240 and h = 360 
the Bitcoin series gives MAPE
values of 0.00835 and 0.010 
for linear regression and values of 0.0072 and 
0.0081 for univariate lstm respectively).

These results show how such innovative partitioning approach allowed us to avoid the ``random walk problem'', finding that best results are obtained using more than one previous price. Furthermore this method leads to a significant improvement in predictions.
It is worth noting that, from this analysis the best result arise from the Bitcoin series, with a \textit{MAPE} error of $0,007$, a temporal window $k_p$ of $4$ and a translation step $h$ of $120$, obtained with the \textit{LSTM} model. 

\begin{table}[h!]
\caption{LR (left) and MLR (right) results with time regimes}
\label{table:tableEight}
\begin{center}
\begin{tabular}{ |p{1.1cm}|p{1.4cm}|p{1.5cm}|p{0.3cm}|p{0.5cm}|p{1.4cm}|p{1.5cm}|p{0.4cm}|p{0.4cm}|p{0.5cm}|} 
\hline
  \hline
  \textbf{Series} & \textbf{MAPE} & \textbf{rRMSE} & \boldmath{$k_p$} & \textbf{h} & \textbf{MAPE} & \textbf{rRMSE} & \boldmath{$k_p$} & \boldmath{$k_v$} & \textbf{h} \\
  \hline 
  BTC & 0,008 & 0,011 & 5 & 120 & 0,008 & 0,012 & 5 & 1 & 120\\
  MSFT & 0,008 & 0,011 & 2 & 720 & 0,008 & 0,011 & 1 & 1 & 720\\
  INTC & 0,010 & 0,015 & 1 & 360 & 0,010 & 0,015 & 1 & 1 & 360\\
  NKSH & 0,010 & 0,013 & 9 & 480 & 0,010 & 0,013 & 9 & 1 & 480\\
  \hline
\end{tabular}
\end{center}
\end{table}

\begin{table}[h!]
\caption{Univariate (left) and Multivariate LSTM (right) results with time regimes}
\label{table:tableNine}
\begin{center}
\begin{tabular}{ |p{1.2cm}|p{1.3cm}|p{1.6cm}|p{0.3cm}|p{0.5cm}|p{1.3cm}|p{1.6cm}|p{0.3cm}|p{0.3cm}|p{0.5cm}|} 
\hline
  \hline
  \textbf{Series} & \textbf{MAPE} & \textbf{rRMSE} & \boldmath{$k_p$} & \textbf{h} & \textbf{MAPE} & \textbf{rRMSE} & \boldmath{$k_p$} & \boldmath{$k_v$} & \textbf{h} \\
  \hline 
  BTC & 0,007 & 0,012 & 4 & 120 & 0,007 & 0,012 & 1 & 1 & 120\\
  MSFT & 0,008 & 0,011 & 4 & 720 & 0,008 & 0,016 & 4 & 4 & 720\\
  INTC & 0,011 & 0,015 & 8 & 360 & 0,024 & 0,028 & 8 & 8 & 360\\
  NKSH & 0,010 & 0,014 & 1 & 480 & 0,011 & 0,014 & 1 & 1 & 480\\
  \hline
\end{tabular}
\end{center}
\end{table}

\begin{table}[h!]
\caption{Best Benchmarks Results}
\label{table:tableTen}
\begin{center}
\begin{tabular}{ |p{2cm}|p{2.5cm}|p{2.5cm}|p{1.5cm}| } 
\hline
\textbf{Reference} & \textbf{Series} & \textbf{Model} & \textbf{MAPE} \\
  \hline
  \cite{O} & BTC & SVM:0.9-1(Relief) & 0,011 \\
  \hline
  \cite{N} & S\&P BSE SENSEX & SVR & 0,009 \\
  \hline
  {\cite{G}} & MSFT & SVR-CFA & 0,052 \\
  & INTC & SVR-CFA & 0,045 \\ 
  & NKSH & SVR-CFA & 0,046 \\ 
  \hline
\end{tabular}
\end{center} 
\end{table}

Another interesting consideration that arises from the results is that, as stated previously in the analysis of the series in their entirety, the linear regression models generally outperform the neural networks ones, while in the short-time regimes approach best results were obtained with the \textit{LSTM} network.

For a direct feedback with our results, shown in tables \ref{table:tableFour}, \ref{table:tableFive}, \ref{table:tableEight} and \ref{table:tableNine}, we report in table~\ref{table:tableTen} the best results obtained in the papers to which we compared to. It is noticeable that our results outperform those obtained in the benchmark papers.

\section{Conclusion}
The results, obtained considering the series in their totality, reflect the considerations made in the introduction of this paper, namely the predictions of the Bitcoin closing price series are worse, in terms of $MAPE$ error, than those obtained for the benchmark series (Intel, Microsoft and National Bankshares). This is probably due to at least two reasons: high volatility of the Bitcoin price and market immaturity for cryptocurrencies. This is confirmed by the statistics reported in tables \ref{table:tableOne} and \ref{table:tableTwo}.

The results obtained partitioning the dataset into shorter sequences also confirmed the kindness of our hypothesis of identifying time regimes that do not resemble a random walk and that are easier to model, finding that best results are obtained using more than one previous price.
It is worth noting that, with this novel approach we obtained the best results for the Bitcoin price series, rather than for the stock market series as happened in the analysis of the series in their totality. As stated before, this is probably due to the high volatility of the Bitcoin price, in fact it is no accident that the best result was found for the time regime identified by a translation step $h$ of $120$, where the Bitcoin prices are more distributed around the mean, showing a lower variance. This is confirmed by the standard deviation values shown in table~\ref{table:tableTwo}.

It is important to emphasize that the innovative approach proposed in this paper, namely the identification of short-time regimes within the entire series, allowed us to obtain leading-edge results in the field of financial series forecasting.

Comparing our best result with those obtained in the considered benchmark papers, 
our result represents one of the best found in the literature.
We highlight that we obtained, both for the Bitcoin and the traditional market series, better results than the benchmark ones. Precisely, for Bitcoin we obtained a \textit{MAPE} error of $0,007$, while the benchmark best one \cite{O} is $0,011$.
For the stock market series our algorithms outperform those of benchmarks even more. In fact, our errors are as low as between $15\%$ and $30\%$ with respect to the reference errors reported in the literature.

As regards the implemented algorithms, our best result was found with the \textit{LSTM} network, but from the point of view of execution speed the linear regression models outperform neural networks.












\bibliographystyle{plain}

\end{document}